\documentclass[preprint,12pt]{elsarticle}

\usepackage{color}
\usepackage{graphicx}
\usepackage{amssymb}
\usepackage{amsfonts}
\usepackage{amsmath}
\usepackage{mdframed}

\usepackage{xspace}
\usepackage{fancyvrb}
\usepackage{placeins}
\usepackage{longtable,multirow}
\usepackage{hyperref}
\usepackage{url}

\usepackage{blindtext}
\usepackage{enumitem}

\usepackage[toc,page]{appendix}
\usepackage{lineno}

\newcounter{bla}

\newcommand{\smodelsnn}{\textsc{SModelS}}
\newcommand{\smo}{\textsc{SModelS}}
\newcommand{\smoold}{\textsc{SModelS}\,v1.1}
\newcommand{\smonew}{\textsc{SModelS}\,v1.2}
\newcommand{\smoversion}{\textsc{SModelS}\,v1.2.2}
\newcommand{\Ztwo}{$\mathbb{Z}_2$}

\def\code#1{{\tt{#1}}}

\def\CL{{\rm{CL}}}

\newenvironment{optionlist}[1]                                                        
{\begin{list}{}                                                                       
  {\setlength{\labelwidth}{#1}                                                        
   \setlength{\rightmargin}{0cm}                                                      
   \setlength{\leftmargin}{\rightmargin}                                              
   \addtolength{\leftmargin}{\labelwidth}                                             
   \addtolength{\leftmargin}{\labelsep}                                               
   }                                       
}{\end{list}}

\journal{Computer Physics Communications}

\begin{document}

\begin{frontmatter}

\title{SModelS v1.2: long-lived particles, combination of signal regions, and other novelties}

\author[a]{Federico Ambrogi}
\author[b]{Juhi Dutta}
\author[c]{Jan Heisig}
\author[d]{Sabine Kraml}
\author[e]{Suchita Kulkarni}
\author[f,g]{Ursula Laa}
\author[h]{Andre Lessa}
\author[e]{Philipp Neuhuber} 
\author[d]{Humberto Reyes-Gonz\'alez}
\author[e]{Wolfgang Waltenberger}
\author[e]{Matthias Wolf}

\address[a]{University of Vienna, Faculty of Physics, Boltzmanngasse 5, A-1090 Wien, Austria}
\address[b]{Regional Centre for Accelerator-based Particle Physics, Harish-Chandra Research Institute, 
HBNI, Chhatnag Road, Jhusi, Allahabad-211019, India}
\address[c]{Centre for Cosmology, Particle Physics and Phenomenology (CP3), Universit\'e catholique de Louvain, Chemin du Cyclotron 2, B-1348 Louvain-la-Neuve, Belgium}
\address[d]{Laboratoire de Physique Subatomique et de Cosmologie, Universit\'e
  Grenoble-Alpes, CNRS/IN2P3, 53 Avenue des Martyrs, F-38026 Grenoble, France}
\address[e]{Institut f\"ur Hochenergiephysik,  \"Osterreichische Akademie der
  Wissenschaften, Nikolsdorfer Gasse 18, 1050 Wien, Austria}
\address[f]{School of Physics and Astronomy, Monash University, Clayton, VIC 3800, Australia}
\address[g]{Department of Econometrics and Business Statistics, Monash University, Clayton, VIC 3800, Australia}
\address[h]{Centro de Ci\^encias Naturais e Humanas, Universidade Federal do ABC, Santo Andr\'e, 09210-580 SP, Brazil}

\begin{abstract} 
\smo\ is an automatised tool enabling the fast interpretation of simplified model results from the LHC within any model of new physics 
respecting a \Ztwo\ symmetry. With the version 1.2 we announce several new features. First, previous versions were restricted to missing energy signatures and assumed prompt decays within each decay chain. \smo v1.2 considers the lifetime of each \Ztwo-odd particle and appropriately takes into account missing energy, heavy stable charge particle and R-hadron signatures. Second, the current version allows for a combination of signal regions in efficiency map results whenever a covariance matrix is available from the experiment. This is an important step towards fully exploiting the constraining power of efficiency map results.   
Several other improvements increase the user-friendliness, such as the use of wildcards in the selection of experimental results, and a faster database which can be given as a URL.  
Finally, smodelsTools provides an interactive plots maker to conveniently visualize the results of a model scan. 
\end{abstract}


\end{frontmatter}


{\bf NEW VERSION PROGRAM SUMMARY}

\begin{small}
\noindent
{\em Program Title:} SModelS                                   \\
{\em Licensing provisions:} GPLv3                              \\
{\em Programming language:} Python3                         \\
{\em Journal reference of previous version:} Comput.\ Phys.\ Commun.\  {\bf 227} (2018) 72 \\
{\em Does the new version supersede the previous version?:} Yes  \\
{\em Reasons for the new version:} Addition of new features\\
 
{\em Summary of revisions:}\\
The most important new features in v1.2 are the combination of signal regions in efficiency map results whenever a 
covariance matrix is available from the experiment, and the implementation of heavy stable charge particle and R-hadron signatures.  
Moreover, the database of experimental results can now 
be given as a URL, and the pickling has been improved to make the database faster. Other improvements include 
that wildcards are allowed when selecting analyses, datasets or topologies, and that the path to the model file, 
formerly required to be smodels/sparticles.py, can be specified in the parameters card.  
For the convenience of the user, we also provide a tool to make interactive plots to visualize results of a model scan.  
Finally, the whole code now also runs with Python\,3, which has become the
recommended default, and it can now be installed in its source directory. 

{\em Nature of problem:}\\  
The results for searches for new physics beyond the Standard Model (BSM) at the Large Hadron Collider are often communicated by the experimental collaborations in terms of constraints on so-called simplified models spectra (SMS). 
Understanding how SMS constraints impact a realistic new physics model, where possibly a multitude of production channels and decay modes are relevant, is a non-trivial task. 

{\em Solution method:}\\  
We exploit the notion of simplified models to constrain full models by ``decomposing'' them into their SMS components. 
A database of SMS results obtained from the official results of the ATLAS and CMS collaborations, but in part also from `recasting' the experimental analyses,  can be matched against the decomposed model, resulting in a statement to what extent the model at hand is in agreement or contradiction with the experimental results. Further useful information on, 
e.g., the coverage of the model's signatures  is also provided.

{\em Additional comments including Restrictions and Unusual features:}\\ 
At present, only models with a \Ztwo-like symmetry can be tested. Each SMS is defined purely by the vertex structure and the final-state particles; initial and intermediate BSM particles are described only by their masses, production cross sections, branching ratios and total width. Possible differences in signal selection efficiencies arising, e.g., from different production mechanisms or from the spin of the BSM particles, are ignored in this approach. Since only part of the full model can be constrained by SMS results, \smo\ will always remain more conservative (though orders of magnitude faster) than ``full recasting'' approaches. \\

\end{small}

\clearpage

\section{Introduction}\label{intro}

\smo\ is an automatic, public tool for interpreting simplified-model results from the LHC. It is based on a general procedure to decompose Beyond the Standard Model (BSM) collider signatures respecting a \Ztwo\ symmetry into Simplified Model Spectrum (SMS) topologies. Our method provides a way to cast BSM predictions for the LHC in a model independent framework, which can be directly confronted with the relevant experimental constraints.

In Ref.~\cite{Ambrogi:2017neo}, we introduced the use of efficiency maps%
\footnote{By efficiency maps we mean maps of acceptance$\times$efficiency (${\rm eff} = A\times\epsilon$).} 
in SModelS. The advantage of efficiency maps (EMs) over cross section upper limit (UL) maps is that they 
allow for combining contributions from different simplified-model topologies to the same signal region 
(SR) of a given experimental analysis.  
This can significantly enhance the coverage and constraining power of the simplified model results. 
However, a drawback compared to UL-type results so far was that only the most sensitive SR could 
be used in EM-type results. 

The next step further is to statistically combine the contributions to the different SRs within an analysis. 
This is possible if a) the SRs are exclusive, i.e.\ non-overlapping, and b) a likelihood model describing the background 
correlations is available. To this end, the CMS collaboration has recently started to provide covariance matrices 
within the simplified likelihood framework~\cite{simplifiedlikelihoods} for their supersymmetry (SUSY)
results. The support for these covariance matrices and hence the combination of signal regions 
is one of the main new features presented here. 

The other major extension presented in this paper concerns the kind of topologies which can be treated by \smo: so far, in previous versions, only BSM searches in final states with missing transverse momentum (MET) could be taken into account. Furthermore, all BSM states were required
to have prompt decays or be stable (at collider scales).
However, well-motivated BSM theories also often feature 
meta-stable states or long-lived particles (LLPs) which are
electrically or color charged.
\smonew\ now supports two important classes of LLP signatures: heavy stable
charged particles (HSCPs) and R-hadron signatures. 
To this end, the decomposition procedure was extended to include a {\tt final state} property, which corresponds to the distinct types of signatures: MET, HSCP or R-hadron.
The inclusion of LLP signatures in \smo\ and its physics impact were first presented in \cite{Heisig:2018kfq}; here 
give some more technical details of the implementation and how to use it.  \\

\noindent
Other novelties in this new version include:

\begin{description}[noitemsep]
\item[\, --] the database of experimental results can now be given as a URL; 
\item[\, --] new experimental results were added to the database; 
\item[\, --] the pickling has been improved to make the database faster; 
\item[\, --] wildcards are allowed in the selection of analyses, topologies or datasets; 
\item[\, --] the path to the model file, formerly required to be smodels/particles.py, can be specified 
in the parameters file (parameters.ini);
\item[\, --] more detailed information is available in the default output formats;  
\item[\, --] a tool to make interactive plots was added to smodelsTools;  
\item[\, --] the default python version changed from Python\,2 to Python\,3;
\item[\, --] a makefile is provided to install \smo\ in the source folder, as an alternative to the previous installation in the default system or user directory. 
\end{description}

The concrete version we are presenting in the following is \smoversion. 
In section~\ref{parameterfile} we first explain the new features accessible via the parameters file 
before discussing in more detail the combination of signal regions in section~\ref{covariances}.  
The implementation of LLP signatures is presented in section~\ref{longlived}. 
The module in smodelTools to make interactive plots is presented in section~\ref{iplots}. 
Further updates of the code and the database are documented in section~\ref{others}, 
and the new installation procedure in section~\ref{installation}. 
We conclude in section~\ref{conclusions}.

\clearpage
\section{New features accessible via the parameters file}\label{parameterfile}

The basic options and parameters used by \code{runSModelS.py} are defined in the
parameter file. An example parameter file, including all available parameters
together with a short description, is given in \code{parameters.ini} in the \smodelsnn\ distribution. 
If no parameter file is specified, the default parameters stored in
\code{smodels/etc/\allowbreak parameters\_default.ini} are used. 
Below we describe the parts which are new with respect to \smoold~\cite{Ambrogi:2017neo}.

\noindent\begin{description}
\item[\emph{Section ``options''}:] main options for turning SModelS features on or off.
\end{description}

\begin{itemize}
  
  \item {}
  \textbf{combineSRs} (True/False): Set True to combine signal regions when a covariance matrix is available. 
  Note that this might take a few seconds per point. False uses only the most sensitive (best expected) SR. 
  
  The combination of SRs is described in more detail in section~\ref{covariances} below.
 
\end{itemize}

\noindent\begin{description}
\item[\emph{Section ``particles''}:] allows to define the BSM particle content. 
\end{description}

\begin{itemize}
  \item {}
  \textbf{model} (path): path to the Python file which describes the 
  $\mathbb{Z}_2$-even and $\mathbb{Z}_2$-odd particle content of the BSM model
to be tested.\footnote{Formerly the particle content had to be specified in
the hard-coded file path \code{smodels/particles.py}.} 
  This can be given either in Unix file notation, e.g.
  
   \quad  \code{model=share/models/mymodel.py} 

   or as a Python module path, e.g.

   \quad \code{model=share.models.mymodel}  

If \code{model=mymodel}, the file \code{mymodel.py} is searched for in the current 
working directory as well as in \code{smodels/share/models/}. The latter directory contains a  few sample model files.
If no model is specified, the standard MSSM particle content is taken as the default. 
 
Note that from v1.2 onwards the quantum numbers of each BSM particle must
also be specified in order to classify long lived particles as missing energy, HSCP or R-hadron final states. The mapping of quantum numbers to final states is defined in 
\code{smodels/theory/particleNames.py}.

\end{itemize}

\noindent\begin{description}
\item[\emph{Section ``database''}:] allows choice of database version and/or selection of a subset of
experimental results. 
\end{description}

\begin{itemize}

  \item {}
  \textbf{path}: the absolute or relative path to the database. The user can
  supply either the directory name of the database, or the path to the 
  pickle file. From v1.1.3 onwards, it is also possible to provide the URL to the  
  official database release, e.g.: 
  
  \code{path = http://smodels.hephy.at/database/official122}\,.

  The \code{official122.pcl} file has about 400 MB; downloading it is often faster than 
  parsing the text database oneself.   
  Moreover, it has the advantage of avoiding any machine dependence in the pickling.  
  The list of database releases is available at \cite{database:releases}. Note that 
  separate URLs are provided for database versions which do not or do include the 8~TeV efficiency maps from Fastlim-1.0~\cite{Papucci:2014rja}. Furthermore, some
  pickle files can only be used with the corresponding SModelS version. 
  
	Note: from v1.2.2 onwards, the shorthand notation \code{path=official} (or
  \code{path=official\_fastlim}) points to the official database 
  without (with) Fastlim of the same version as the SModelS code that is being used.
    
  \item {}
  \textbf{analyses} (list of results): set to {\emph{all}} to use all available results.
  If a list of experimental analyses is given, only these will be used. 
  
  Wildcards (*, ?, [$\langle$list-of-or’ed-letters$\rangle$]) are expanded in
the same way a Unix shell does wildcard expansion for file names.  
  For example, ATLAS* will use all ATLAS results, while CMS-SUS-16* will use only published CMS SUSY 
  results with an ID from 2016.   
  
  To filter by centre-of-mass energy, use a suffix beginning with a colon, like :13*TeV 
  (note that the asterisk in the suffix is a multiplication operator and not a wildcard). This can be combined with wildcards, e.g.,  
  CMS*:13*TeV will use only CMS Run~2 results. 
    
  \item {}
  \textbf{txnames} (list of topologies): set to {\emph{all}} to use all available
  simplified model topologies.  The topologies are labeled
  according to the TxName convention~\cite{smodels:dictionary}.  
  If a list of TxNames is given, only the corresponding topologies will be considered.
  Wildcards (*, ?, [$\langle$list-of-or’ed-letters$\rangle$]) can now also be
used. As for analyses above, they are expanded in the same way the shell does
wildcard expansion for file names.  For example, txnames = T[12]*bb* picks all
txnames beginning with T1 or T2 and containing bb. As of the time of writing
these were: T1bbbb, T1bbbt, T1bbqq, T1bbtt, T2bb, T2bbWW, T2bbWWoff.   
  
  \item {}
  \textbf{dataselector} (list of datasets): set to {\emph{all}} to use all available
  data sets. If dataselector = upperLimit (efficiencyMap), only UL-type results
  (EM-type results) will be used. Furthermore, if a list of signal regions
  (data sets) is given, only the experimental results
  containing these datasets will be used.  Wildcards can be used in the same way as for analyses and txnames.  
  
\end{itemize}

\noindent\begin{description}
\item[\emph{Section ``python/xml printers''}:] options for printing the results.
\end{description}

\begin{itemize}
  
  \item {}
  \textbf{addTxWeights} (True/False): set to True to show individual
contribution from each topology to the total theory prediction for EM-type results. 
  
\end{itemize}

\section{Combination of signal regions}\label{covariances}

\subsection{Method}

If the experiment provides a covariance matrix together with efficiency maps  for SMS topologies, 
the signal contributions in different signal regions can be combined~\cite{simplifiedlikelihoods,Buckley:2018vdr}.%
\footnote{We stress here that, with or without covariances, EMs are needed {\it for all signal regions}. 
Since the best SR can change depending on the model being tested, a single map containing the efficiencies 
for the best SR in each bin cannot be used in a general way.} 
This is implemented in SModelS v1.1.3 onwards following the (symmetric) simplified likelihood approach of~\cite{simplifiedlikelihoods}, 
and easily extendible to the more general treatment described in~\cite{Buckley:2018vdr} once experimental results are available in that format.

SModelS allows for a marginalisation as well as a profiling of the nuisances, with profiling being the default.  
As CPU performance is a concern in SModelS, we try to aggregate the official results, which can comprise a very large number of signal regions, 
to an acceptable number of aggregate regions. Here ``acceptable'' means as few
aggregate regions as possible without significant loss in precision or
constraining power. The CPU time required scales roughly linearly with the
number of signal regions, so aggregating e.g.\ from 80 to 20 signal regions
means gaining a factor of four in computing time.

For the computation of the 95\% confidence level from the likelihoods, a $\CL_s (=\CL_{sb}/\CL_{b})$ limit~\cite{Read:2002hq} is computed from the test statistic $q_\mu$, as described in section 2.4, Eq.~(14), in \cite{Cowan:2010js}. We then search for ${\CL_s} = 0.95$ using Brent's bracketing technique through the SciPy optimize library, see~\cite{scipy:optimize:brentq}.

When using \code{runSModelS.py}, the combination of signal regions is turned on or off with the parameter combineSRs described in the previous section. Per default, combineSRs=False, in which case only the result from the best expected signal region (best SR) is reported. 
If combineSRs=True, both the combined result and the result from the best SR are quoted. 
If the user writes his/her own python code, the combination of SRs is envoked by setting combinedResults=True in \code{theoryPredictionsFor()}. In the same instance, one can switch between profiling and mariginalizing with marginalize=True/False, False being the default. An explicit example is discussed in the ``How To's'' in the online manual~\cite{manual:online}.

\subsection{Available results in the database}

The CMS SUSY group is providing covariance matrices for most of their analyses, so far under the assumption of Gaussian errors.  
However, only two CMS analyses also provide simplified-model efficiency maps for each SR: 
CMS-SUS-16-050~\cite{Sirunyan:2017pjw} and CMS-PAS-SUS-16-052~\cite{CMS-PAS-SUS-16-052}.\footnote{%
Other CMS Run~2 SUSY analyses provide covariance matrices but no efficiency maps. 
ATLAS analyses typically give efficiencies for the best SR only; moreover they do not provide covariances.} 
Since the former analysis has significantly non-Gaussian background uncertainties, as we have checked, the covariance matrix provided for it is not a good approximation~\cite{Buckley:2018vdr}. 
For this reason, at present only the covariance matrix for CMS-PAS-SUS-16-052 is included in the \smo\ database. 

Concretely, CMS-PAS-SUS-16-052 is a search for supersymmetry with compressed mass spectra in events with at least
one soft lepton, moderate to high values of missing transverse momentum
$p_{\rm T}^{\rm miss}$, and one or two hard jets, compatible with the
emission of initial-state radiation.  
It targets scenarios of stop-pair production, $pp\to\tilde t_1^{}\tilde t_1^*$, where the mass difference to the $\tilde\chi^0_1$  
is smaller than the mass of the $W$ boson. 
The analysis has 44 SRs. The simplified model interpretations assume either four-body stop decays, 
$\tilde t\to bf\bar f'\tilde\chi^0_1$ (T2bbWWoff) or decays via an intermediate chargino, 
$\tilde t\to b \tilde\chi^+_1$, $\tilde\chi^+_1\to f\bar f'\tilde\chi^0_1$ (T6bbWWoff).  
Efficiency maps for all 44 SRs are provided for both simplified models. 

\begin{figure}[ht!]\centering
\hspace*{-6mm}\includegraphics[width=0.56\textwidth]{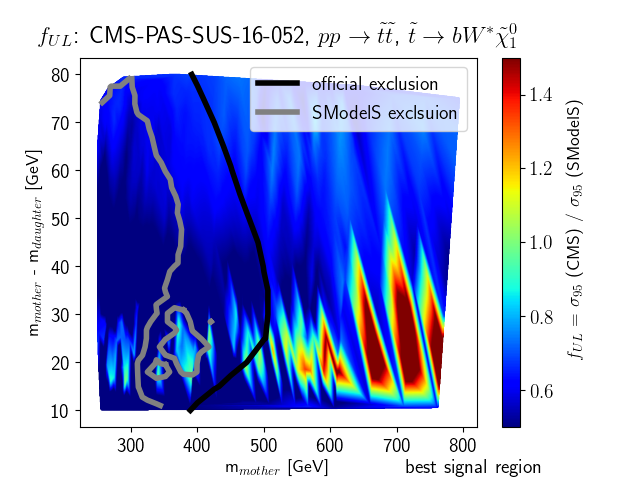}\hspace*{-4mm}\includegraphics[width=0.56\textwidth]{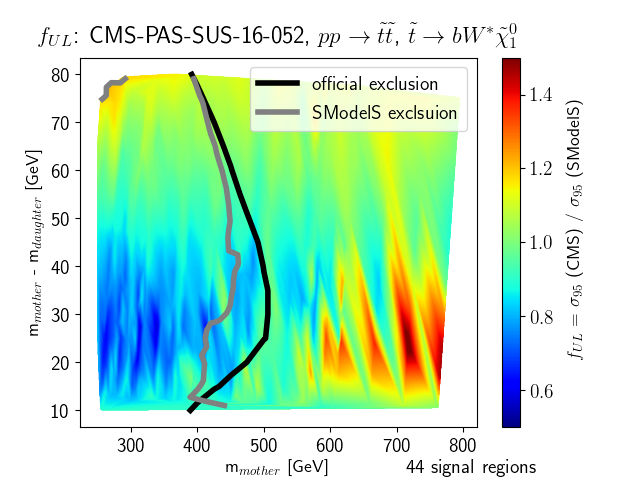}\\
\includegraphics[width=0.56\textwidth]{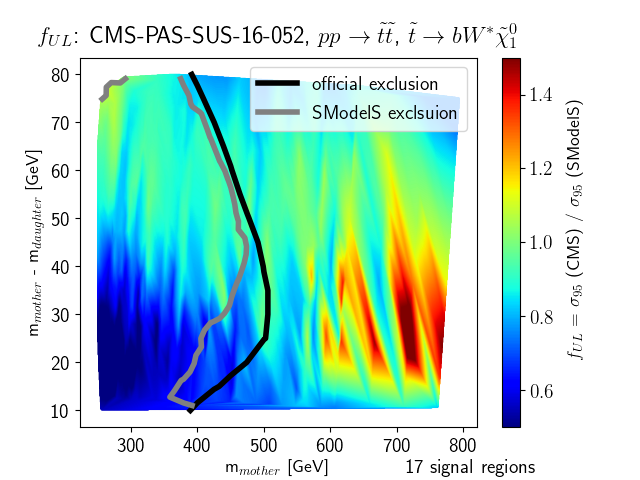}
\caption{\label{fig:combineSRvalidation} Comparison of exclusion curves from
CMS-PAS-SUS-16-052 efficiency maps for the T2bbWWoff simplified model, using
only the best signal region (top left), the combination of all 44 signal regions
(top right) and the combination of 17 aggregate signal regions (bottom). The
colour map shows the $f_{\rm UL}$-value obtained from dividing the official limit
by the one obtained with SModelS. The grey lines indicate the SModelS exclusion $r=1$.
For comparison, the black line shows the official CMS exclusion curve.}
\end{figure}

Figure~\ref{fig:combineSRvalidation} shows validation plots for the T2bbWWoff simplified model, comparing the 95\% CL cross section upper limit derived by SModelS to the official UL from CMS across the SMS mass plane for three cases: using only the best SR (top left), combining all 44 SRs (top right), and aggregating to 17 SRs before combination (bottom). 
The $x$ and $y$ axes are $m_{\tilde t}$ and $m_{\tilde
t}-m_{\tilde\chi^0_1}\equiv \Delta m$, respectively, and the colour code shows
$f_{\rm UL}\equiv \sigma^{ul}_{95}({\rm CMS})/\sigma^{ul}_{95}({\rm SModelS})$. A value of $f_{\rm UL}=0.8$ means that our SR combination  
gives a 20\% weaker limit than the official CMS result, a value of $f_{\rm UL}=1.2$ means that it is 20\% too aggressive.
Also shown are the official exclusion line (black) and the exclusion line derived from SModelS (grey). 

We see that using only the best SR  leads to too weak a limit: the mean $f_{\rm UL}$ is about $0.6$ and the mass limit is up to about 150~GeV too low---compared to the maximum reach of $m_{\tilde t}\approx 500$~GeV this is significant. 
In contrast, the combination of all 44 SR performs much better for reproducing the official CMS result. 
Aggregating the 44 original SRs to 17 speeds up the calculation by more than a factor 2 while still giving a result very close to the full combination. 
The same holds true for the T6bbWWoff topology. 
We therefore included the efficiency maps for 17 aggregated SRs as CMS-PAS-SUS-16-052-agg in the \smo\ database. 
The covariance matrix is given in the \code{globalInfo.txt} file in the same folder. 
Combining 17 SRs takes less than 2 sec on an average 4-core Intel i5 desktop computer.

\section{Implementation of heavy stable charge particle and R-hadron signatures}\label{longlived}

The implementation of HSCP and R-hadron signatures required several important changes in the \smo\ code as detailed below. 
For the user this is primarily noticeable in the ``missing topologies'' output.  
The usage in particular of \code{runSModelS.py} remains the same as before.   

\subsection{Extension of decomposition procedure}

Taking the BSM particle masses, quantum numbers, total decay widths, 
branching ratios (BRs) and total production cross-sections as input, \smo\ 
performs a decomposition of the collider signature of any \Ztwo\ symmetric 
BSM model into a coherent sum of simplified-model topologies. The resulting 
topologies exhibit a two-branch structure emerging from the 
production of two \Ztwo-odd BSM states and their subsequent cascade decays. 
In the v1.0 and v1.1 releases, it was assumed that all BSM decays are prompt 
and that all cascade decays end in a stable neutral BSM particle leading to 
MET final states (otherwise, if {\tt checkInput=True}, an error was issued). 
From v1.2 onwards this has been generalized in two ways. 

First, instead of always assuming a MET final state, 
we allow each cascade decay to terminate in any BSM particle.
Using the quantum numbers defined for the BSM particle (see section~\ref{parameterfile}), the final state is then classified 
as a MET signature, an HSCP or an R-hadron. 
If it does not fall in any of these categories, an error message is displayed.

\begin{figure}[t]
\centering
\hspace*{-5mm}\includegraphics[width=0.98\textwidth]{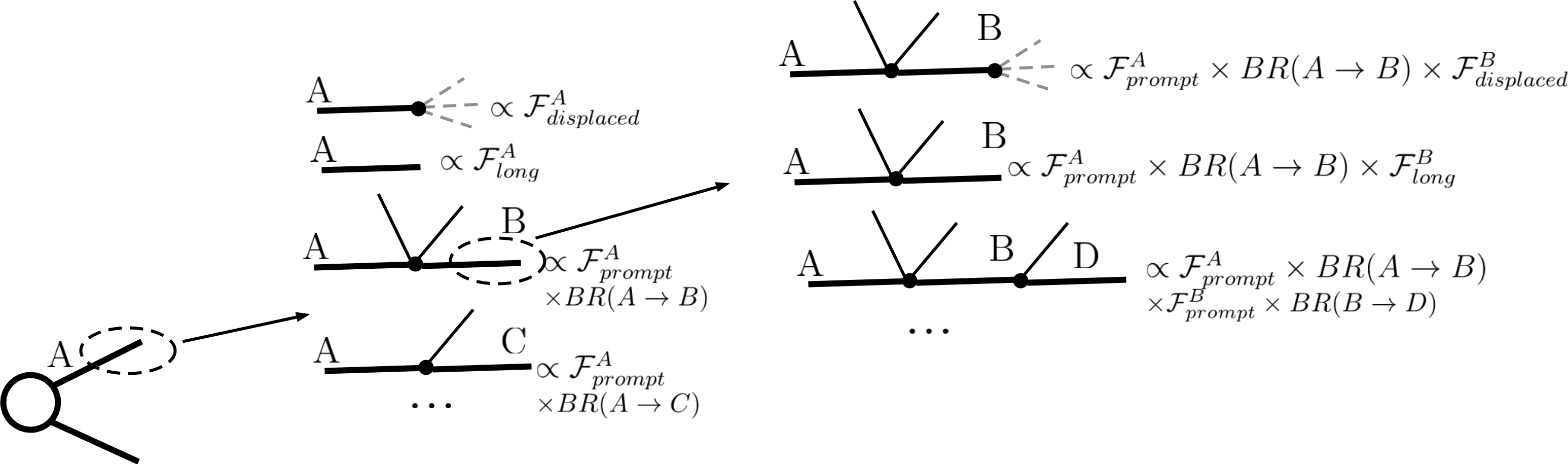}
\caption{Decomposition into simplified model topologies and computation of weights in \smo\,v1.2.}
\label{fig:decomposition}
\end{figure}

Second, during the decomposition into simplified models, at each step in the cascade we compute the probabilities for the respective BSM particle to decay promptly ($\mathcal{F}_\text{prompt}$) and to decay outside the detector ($\mathcal{F}_\text{long}$)~\cite{Heisig:2015yla,Heisig:2018kfq}. 
The procedure is illustrated schematically in Fig.~\ref{fig:decomposition}. 
The fraction of decays which take place inside the detector (labeled simply as ``displaced decays'') is approximated 
as $\mathcal{F}_\text{displaced} = 1 - \mathcal{F}_\text{long} - \mathcal{F}_\text{prompt}$.
Note that the final states appearing in the displaced vertex are not stored during decomposition, since \smo\ is currently 
not able constrain displaced decay signatures; this is left for future work. 
As a result, for the time being, all elements with displaced decays will be identified as missing topologies.

Concretely,  $\mathcal{F}_\text{long}$ and $\mathcal{F}_\text{prompt}$ are computed from the respective particle
proper lifetime, $\tau$, using the approximation
\begin{equation}
\label{eq:fprompt}
\mathcal{F}_\text{prompt} = 1-\exp\left(-\frac{1}{c\tau}\left\langle\!\frac{\ell_\text{inner}}{\gamma \beta} \!\right\rangle_{\!\!\text{eff}\,}\right)
\end{equation}
and
\begin{equation}
\label{eq:Flong}
\mathcal{F}_\text{long} = \exp\left(-\frac{1}{c\tau}\left\langle\!\frac{\ell_\text{outer}}{\gamma \beta} \!\right\rangle_{\!\!\text{eff}\,}\right)\,.
\end{equation}
We choose $\langle\ell_\text{inner}/\gamma\beta\rangle_\text{eff}=1\,$mm and $\langle\ell_\text{outer}/\gamma\beta\rangle_\text{eff}=7\,$m,\footnote{As we currently include CMS results only, $\ell_\text{outer}$ corresponds to the CMS detector size. For the inclusion of ATLAS results $\langle\ell_\text{outer}/\gamma\beta\rangle_\text{eff}$ would have to be adjusted accordingly.} which provides a good approximation to the result of a full simulation as shown in~Appendix~B of~\cite{Heisig:2018kfq}. 

After decomposition the respective weight $\tilde{\sigma}$ of each simplified-model topology is hence given by 
\begin{equation}
\label{eq:weights}
\tilde{\sigma} = \sigma_\text{prod}\left(\prod_i
\text{BR}_i \times \mathcal{F}_\text{prompt}^i \right)
\mathcal{F}_\text{long/displaced}^X \mathcal{F}_\text{long/displaced}^Y \,,
\end{equation}
where $\sigma_\text{prod}$ is the production cross section of the mother particles 
and $X,Y$ are the \Ztwo-odd final states of the two cascades. The index $i$ runs over all intermediate \Ztwo-odd particles. 
For each chain, the last probability factor is given by $\mathcal{F}_\text{long}$ for decays
which take place outside the detector, or by $\mathcal{F}_\text{displaced}$ for displaced decays.

\subsection{Extension of SModelS bracket notation}

\begin{figure}[t]
\centering
\hspace*{-5mm}\includegraphics[width=0.7\textwidth]{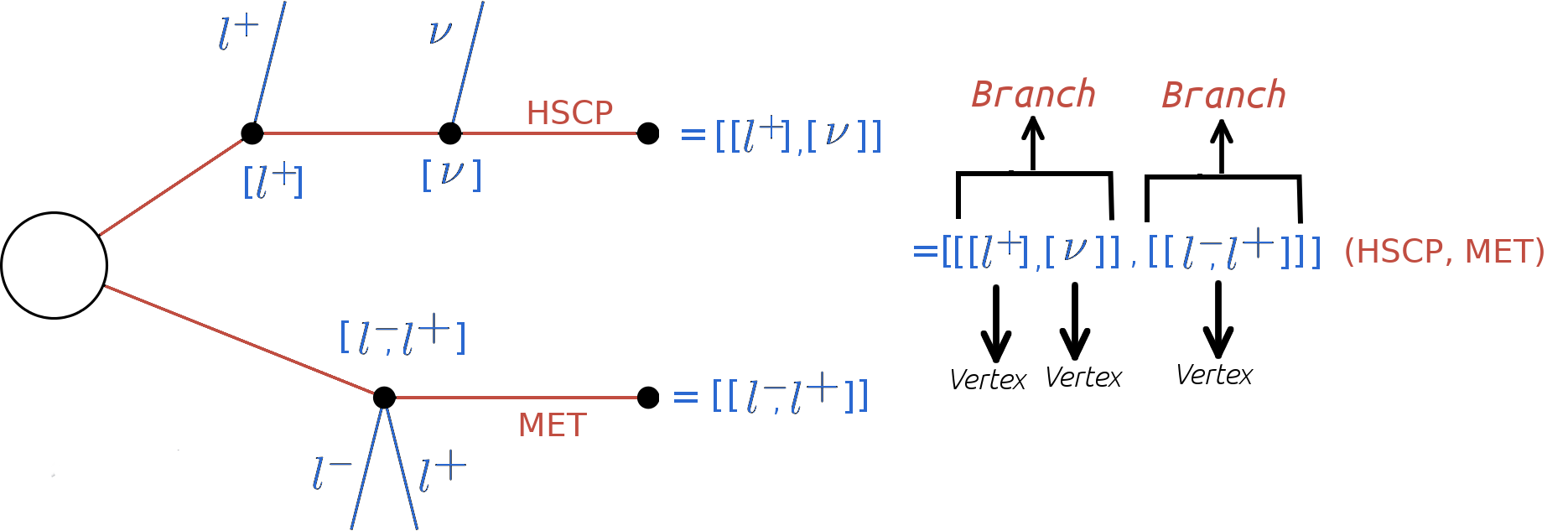}
\caption{Extension of the \smo\ bracket notation to include the \Ztwo-odd final state description, in this example (HSCP, MET).}
\label{fig:bracketNotation}
\end{figure}

As explained in detail in~\cite{Ambrogi:2017neo}, inside \smo\ 
the structure and final states of elements are represented in textual form using a nested brackets notation. In order to fully specify all the information of a given SMS topology (an {\em element}), we must also include the list of masses for the \Ztwo-odd states, the list of \Ztwo-odd final states and the element weight. The masses for the \Ztwo-odd BSM states are represented by a mass array for each branch. 
An example is given in Fig.~\ref{fig:bracketNotation}.

The quantum numbers of this BSM final state are essential for defining which type of signature this element represents. In an element the \Ztwo-odd final state quantum numbers are mapped to final state signatures, as defined in the \code{particleNames} module. 
Currently the defined types of final states are: ‘MET’, ‘HSCP’,
‘RHadronG’, ‘RHadronQ’, the latter two being color-octet and color-singlet
R-hadrons, respectively. New final state types can easily be added in this module.

\subsection{HSCP and R-hadron results in the database}

\begin{figure}[t]
\centering
\hspace*{-5mm}\includegraphics[clip, trim={0.8cm 5cm 1.4cm 3.3cm}, width=0.98\textwidth]{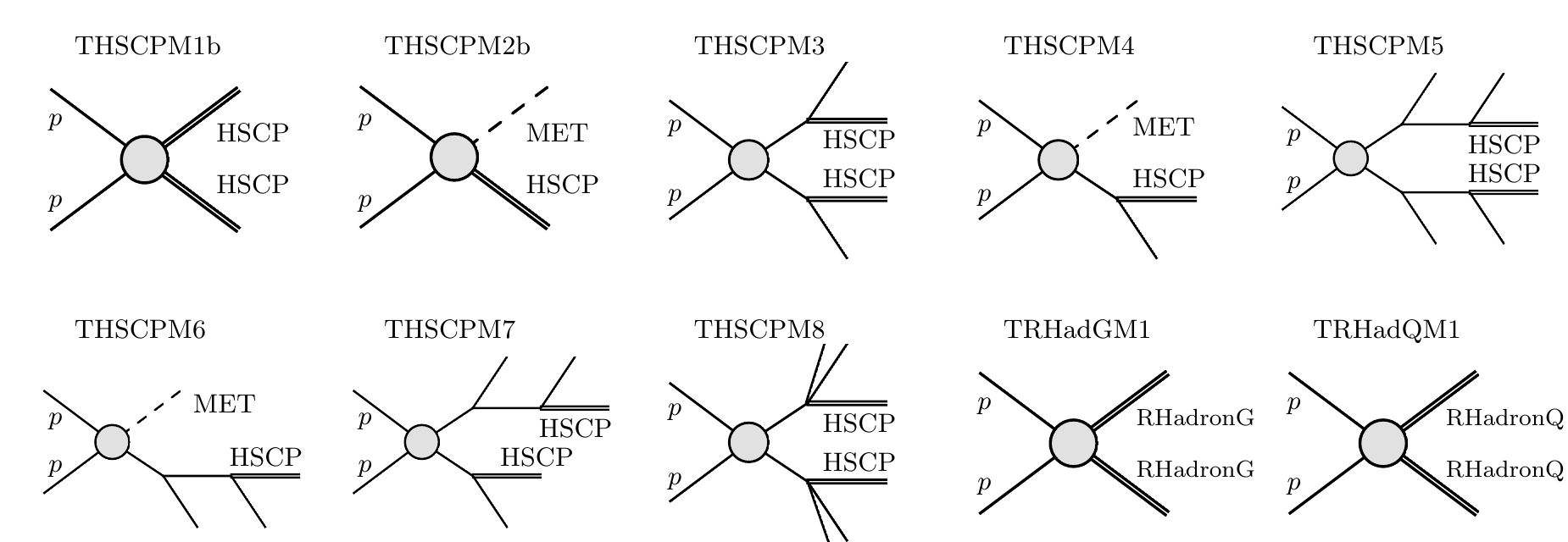}
\caption{Simplified model topologies 
containing HSCP and R-hadron final states 
included in the \smoversion\ database.
A double line represents an HSCP or R-hadron, 
while a dashed line represents an arbitrary
decay chain terminating in a neutral LLP (providing 
a MET signature). The labels of the topologies 
correspond to the naming convention 
used in the database. 
}
\label{fig:txnames}
\end{figure}

The v1.2.2 database includes results from the CMS searches for lepton-like HSCPs (carrying one unit of elementary charge), 
color-octet (gluino-like) R-hadrons and color-triplet (squark-like) R-hadrons at 8~TeV~\cite{Chatrchyan:2013oca} and 13~TeV~\cite{CMS-PAS-EXO-16-036} centre-of-mass energies. 
The relevant topologies and their short-hand notation (``txnames'') in the \smo\ database are summarized in Fig.~\ref{fig:txnames}.

We computed efficiency maps for the eight HSCP simplified-model topologies shown in Fig.~\ref{fig:txnames}. To this end, for the 8~TeV analysis we utilized the recasting provided in Ref.~\cite{Khachatryan:2015lla}. For the 13~TeV analysis we performed a dedicated recasting (see Appendix~A of \cite{Heisig:2018kfq} for details). These efficiency maps are included in the \smoversion\ database.

For the R-hadron searches we consider the direct production 
topologies (TRHadGM1, TRHadQM1) only. As R-hadrons are 
strongly produced, their production via cascade decays is 
assumed to be less relevant and hence not considered here. 
We include the respective cross-section upper limits from 
Refs.~\cite{Chatrchyan:2013oca,CMS-PAS-EXO-16-036}
considering the cloud hadronization model (assuming a 50\% 
probability for gluino-gluon bound state formation).

 \clearpage
\section{Interactive plots maker in smodelsTools}\label{iplots}

For a simple and quick visualisation of results from a scan over input files, we now provide
an ``interactive plots maker'' as part of \code{smodelsTools}.
This tool allows to easily produce interactive plots which relate the SModelS output ({\it in python output format}) with information on the user’s model stored in the SLHA files. It gives 2d plots in the parameter space defined by the user, with additional user-defined information appearing in hover boxes. The output is in html format for viewing in a web browser.  
We stress that the aim is not to make publication-ready plots but simply to facilitate the user's analysis of, e.g., the properties of points in a scan. The usage of the interactive plots tool is:

\begin{Verbatim}[commandchars=\\\{\},frame=lines]
smodelsTools.py interactive-plots [-h] [-p PARAMETERS]
       -f SMODELSFOLDER -s SLHAFOLDER [-o OUTPUTFOLDER]
       [-N NPOINTS] [-v VERBOSITY]
\end{Verbatim}

\begin{description}
\item[{\emph{arguments}:}] \leavevmode
\begin{optionlist}{2cm}
\item [-h, -{-}help]  show this help message and exit.
\item [-p PARAMETERS, -{-}parameters PARAMETERS]
path to the parameters file {[}./iplots\_parameters.py{]}.
\item [-f SMODELSFOLDER, -{-}smodelsFolder SMODELSFOLDER]
path to the smodels folder with the SModelS python
output files.
\item [-s SLHAFOLDER, -{-}slhaFolder SLHAFOLDER]
path to the SLHA folder with the SLHA input files.
\item [-o OUTPUTFOLDER, -{-}outputFolder OUTPUTFOLDER]
path to the output folder, where the plots will be
stored {[}./plots{]}.
\item [-N NPOINTS, -{-}npoints NPOINTS]
How many (randomly selected) points will be included in
the plot. If -1, all points will be read and included
{[}-1{]}.
\item [-v VERBOSITY, -{-}verbosity VERBOSITY]
Verbosity (debug, info, warning, error) {[}info{]}.
\end{optionlist}
\end{description}

\noindent
The default values are given in [\,]. A typical usage example is:

\begin{Verbatim}
./smodelsTools.py interactive-plots  \
    -f inputFiles/scanExample/smodels-output/  \
    -s inputFiles/scanExample/slha -p iplots_parameters.py  \
    -o results/iplots
\end{Verbatim}

The above command will read the SModelS output files (in python format)
from the folder \code{inputFiles/scanExample/smodels-output},
the corresponding SLHA input files from \code{inputFiles/scanExample/slha}
and generate a set of HTML files with the interactive plots, which can
be visualized with a regular web browser.

The settings in {\tt iplots\_parameters.py} include:

\noindent

\noindent\begin{description}
\item[\emph{plot\_title}:] main overall title for your plots, e.g.\ the model name. 
\item[\emph{x and y axes:}:] axis label, SLHA block and PDG code number of the variables you want to plot, in a python dictionary form. 
Example:\footnote{Html notation like {\tt <sub>NAME</sub>} for subscript and {\tt <sup>NAME</sup>} for superscript works for axes labels and in the spectrum hover information, but not in the SModelS hover information.} \begin{Verbatim} 
variable_x = {'m<sub>gluino</sub>': ['MASS', 1000021]}
variable_y = {'m<sub>LSP</sub>': ['MASS', 1000022]}
\end{Verbatim}
\item[\emph{spectrum hover information}:] defines which information from the input SLHA file will appear in the hover box.
The syntax is again a python dictionary.
\end{description}
\begin{itemize}
   \item {\tt slha\_hover\_information}: information from the input SLHA file, e.g. model parameters or masses. Example: \\
   {\tt slha\_hover\_information = \{'m(gluino)’: ['MASS’, 1000021], \\ 'm(chi10)’: ['MASS’, 1000022]\}}. 
   \item {\tt BR\_hover\_information}: defines for which particle(s) to display decay channels and branching ratios. Example: \\
   {\tt BR\_hover\_information = \{'BR(gluino)’: 1000021\}}.\\ 
   The output is written in the form $.25[1000022,1,-1]$,  where the first number (0.25) is the branching ratio, and the numbers in $[\,]$ are the PDG codes of the decay products.
    
   WARNING: Lists of branching ratios can be very long, so the may not fit in the hover box. 
   One can define the number of entries with {\tt BR\_get\_top} (default: ‘all’); e.g.\ {\tt BR\_get\_top = 3} will display only the three largest branching ratios.
   \item {\tt ctau\_hover\_information}: displays the mean decay length in meter for the listed particle(s). Example: \\
   {\tt ctau\_hover\_information = \{'ctau(chi1+)’: 1000024\}}.
\end{itemize}
\begin{description}
\item[\emph{SModelS hover information}:] defines, as a list of keywords, which information to display from the SModelS output. 
Example: \begin{Verbatim} 
smodels_hover_information = ['SmodelS_excluded’, 'r_max’, 
'Tx’, 'Analysis’, 'file’]. 
\end{Verbatim}
The options are:
\begin{itemize}
       \item {\tt SModelS\_status}: prints whether the point is excluded or not by SModelS;
       \item {\tt r\_max}: shows the highest r-value for each parameter point (‘False’ if no experimental result applies);
       \item {\tt chi2}: shows, if available, the $\chi^2$ value associated to the highest r-value (if not, the output is ‘False’);
       \item {\tt Tx}: shows the topology/ies which give the highest r-value;
       \item {\tt Analysis}: shows the experimental analysis from which the strongest constraint (r\_max) comes from;
       \item {\tt MT\_max}: shows the missing topology with the largest cross section (in SModelS bracket notation);
       \item {\tt MT\_max\_xsec}: shows the cross section of MT\_max in fb;
       \item {\tt MT\_total\_xsec}: shows the total missing cross section in fb, i.e.\ the sum of all missing topologies cross sections;
       \item {\tt MT\_long\_xsec}: shows the total missing cross section (in fb) in long cascade decays; 
       \item {\tt MT\_asym\_xsec}: shows the total missing cross section (in fb) in decays with asymmetric branches;
       \item {\tt MT\_outgrid\_xsec}: shows the total missing cross section (in fb) outside the mass grids of the experimental results;
       \item {\tt file}: shows the name of the input spectrum file.
\end{itemize} 
\item[\emph{Choice of plots to make}]:
\begin{itemize}
       \item {\tt plot\_data}: choice of which points to plot; the options are: all, excluded, non-excluded points. Example: \\
       {\tt plot\_data = ['all', 'excluded', 'non-excluded']}.
       
       \item {\tt plot\_list}: which quantities to plot in the $x,\,y$ plane; the same options as for \smo\ hover information apply. Example: \\
       {\tt plot\_list = ['r\_max', 'chi2', 'Tx', 'Analysis', 'MT\_max', \\ 'MT\_max\_xsec', 'MT\_total\_xsec']}.
\end{itemize}
The plotted quantities (r$\_$max, chi2, Tx, Analysis, etc.) and kind of points (all, excluded, non-excluded by \smo) are reflected 
in the interactive plots' filenames. Moreover, a file \code{index.html} is created for a convenient access to all plots.  
\end{description}


\section{Other updates in SModelS}\label{others}

\subsection{Database update}\label{database_update} 

In addition to the aggregated EMs and covariance matrix for CMS-PAS-SUSY-16-052 
explained in section~\ref{covariances}, and the HSCP and R-hadron results explained in section~\ref{longlived}, 
the database was extended in several more ways:  
\begin{itemize}
\item CMS Run~2 results: 
84 new UL maps from 19 different CMS SUSY analyses with 36 fb$^{-1}$ where added in v1.1.2. 
The impact of this update was shown in \cite{Dutta:2018ioj}.
\item ATLAS Run~2 results: 
12 new UL maps from 6 ATLAS SUSY analyses with 36 fb$^{-1}$ were added in v1.2.2, and one existing $3.2$~fb$^{-1}$ analysis was augmented 
with an additional EM result. 
\item Finally, EMs relevant for constraining gluino--squark production (see \cite{Ambrogi:2017lov}) were produced by us 
by recasting ATLAS and CMS 8 TeV multi-jet + MET analyses with MadAnalysis\,5~\cite{Dumont:2014tja}.  
This resulted in 51 new EMs (for 8 TeV) added to the v1.2.2 database. 
Their impact will be analysed in detail in a separate publication~\cite{Federico:TGQ}.  
\end{itemize}

The evolution of the database can also be traced via the wiki pages on the \smo\ server: 
{\tt http://smodels.hephy.at/wiki/ListOfAnalyses<VERSION>} and
{\tt http://smodels.hephy.at/wiki/SmsDictionary<VERSION>} document the state
of the database in version {\tt <VERSION>}, with the dots removed (i.e.\ 122 for
version 1.2.2). Leaving {\tt <VERSION>} empty returns the latest official version.

\subsection{Bug fixes}\label{bugs}

\begin{itemize}
  \item \code{smodelsTools.py} has fixpermissions for system-wide installations via \code{pip3};  
  this allows to install programs used by \code{smodelsTools} (Pythia\,6, Pythia\,8, NLLfast) into 
  system-wide directories (if run with \code{sudo});
  \item fixed small issue with pair production of \Ztwo-even particles;
  \item fixed \code{lastUpdate} bug: the date of the last update of a database entry is now printed correctly; 
  \item fixed a problem in \code{particleNames.py} for non-MSSM models;
  \item fixed small issue when clustering elements with asymmetric masses.
\end{itemize}

\section{Installation}\label{installation}

\smo\ is a Python library that requires Python version 2.6 or later. 
It is fully compatible with Python\,3, which is now the recommended default. 
The external Python dependencies are 
\begin{quote}
unum$>=$4.0.0, numpy$>=$1.13.0, scipy$>=$1.0.0, argparse, \\ requests$>=$2.0.0, docutils$>=$0.3 and pyslha$>=$3.1.0. 
\end{quote}
These tools need not be installed separately, as the \smo\ build system takes care of this. 
In addition, the cross section computer provided by \code{smodelsTools.py} requires a C++ compiler (for Pythia\,8~\cite{Sjostrand:2007gs}) 
and a Fortran compiler (for NLLfast~\cite{nllfast} and Pythia\,6~\cite{Sjostrand:2006za}). 
The current default is that both Pythia\,6 and Pythia\,8 are installed together with NLLfast. 
Finally, the database browser provided by \code{smodelsTools.py} requires IPython, while the interactive plotter 
requires plotly and pandas.

From v1.1.3 onwards, \smo\ can be installed in its source directory. 
After downloading the code from 
\begin{quote}
\url{https://github.com/SModelS/smodels/releases} 
\end{quote}
and extracting it, run
\begin{verbatim}
  make smodels
\end{verbatim}

\noindent
in the distribution's top-level directory. The installation will remove redundant folders, install the required dependencies (using \code{pip install}) and compile Pythia\,8 and NLL-fast. 
If the installation fails with an error that no Fortran compiler was found, try  
\begin{verbatim}
  make FC=<path-to-fortran-compiler> smodels
\end{verbatim}

If the cross section computer is not needed, one can replace \code{smodels}
with \code{smodels\_noexternaltools} in the \code{make} command. In case some
Python libraries cannot be installed automatically, we recommend to install them separately with the user's preferred method. Pythia and NLL-fast can also be compiled separately running \code{make externaltools}. 

For more details, see the online documentation~\cite{manual:online}. 


\section{Conclusions}\label{conclusions}

Version 1.2 of \smo\ presented here features the combination of signal regions 
in efficiency map results whenever a covariance matrix is available from the experiment, 
and the implementation of HSCP and R-hadron signatures. 
The  former is a crucial step towards fully exploiting the constraining power of efficiency map results, 
while the latter considerably extends the types of topologies which can be treated by \smo. 
Several other improvements in the code increase the user-friendliness of the tool. 
These include, for instance, 
the use of wildcards in the selection of analyses, topologies or datasets, 
and a faster database which can be given as a URL. 
Moreover, an interactive plots maker was added to \code{smodelsTools} in order to conveniently visualize results of a model scan. 

The v1.2.2 database includes 2733 individual results for 76 different SMS topologies from a total of 75 ATLAS and CMS analyses. 
This is not counting superseded results.  
In particular new 51 ``home-grown'' EMs for 8~TeV have been added, which were produced by us for better covering scenarios 
where gluino-squark associated production is important. 
For Run~2, the database now comprises 9 UL and 3 EM results  from 10 ATLAS analyses, as well as 
89 UL (of which 3 LLP)  and 10 EM (of which 8 LLP) results from 21 CMS analyses. 
The addition of more ATLAS/CMS Run~2 results is ongoing. 

Understanding how the plethora of LHC searches for new physics, conducted in specific final states, 
impact realistic new physics models, where possibly a multitude of production channels and decay modes are relevant, 
is a non-trivial task, but important for guiding future theoretical and experimental investigations. 
We will continue to further extend and improve \smo\ to contribute to this aim. 

\section*{Acknowledgements}

We thank all \smo\ users for their interest, feedback and encouragement.

J.D. is partially supported by funding available from the Department of Atomic Energy, Government of India, for the Regional Centre for Accelerator-based Particle Physics (RECAPP), Harish-Chandra Research Institute (HRI), and by an INFOSYS scholarship for senior students at the HRI.  She thanks moreover the INFRE-HEPNET (IndoFrench Network on High Energy Physics) of CEFIPRA/IFCPAR (Indo-French Centre for the Promotion of Advanced Research) for support for research visits to LPSC Grenoble. 
J.H.\ acknowledges support from the F.R.S.-FNRS, of which he is a postdoctoral researcher, 
and S.K.\ from the IN2P3 project ``Th\'eorie -- LHCiTools'' and the CNRS-FAPESP collaboration grant PRC275431. 
Su.K.\ is supported by the “New Frontiers” program of the Austrian Academy of Sciences and the FWF under project number V592-N27. 
A.L.\ acknowledges support the Sao Paulo Research Foundation FAPESP projects 2015/20570-1 and 2016/50338-6 and 
H.R.G.\ from the Consejo Nacional de Ciencia y Tecnología, CONACyT, scholarship no.\ 291169.

\clearpage
\section*{References}
\bibliographystyle{elsarticle-num}
\bibliography{references}

\end{document}